\documentclass[aps,twocolumn,superscriptaddress]{revtex4}
\usepackage{graphicx}

\newcommand{\ket}[1] {\left| #1 \right\rangle}

\begin{document}

\title{Landau-Zener-Stueckelberg interferometry with driving fields in the quantum regime}

\author{S. Ashhab}
\affiliation{Qatar Environment and Energy Research Institute, Hamad Bin Khalifa University, Qatar Foundation, Doha, Qatar}

\date{\today}

% July 17, 2016

\begin{abstract}
We analyze the dynamics of a two-level quantum system (TLS) under the influence of a strong sinusoidal driving signal whose origin is the interaction of the two-level system with a quantum field. In this approach the driving field is replaced by a harmonic oscillator that is either strongly coupled to the TLS or populated with a large number of photons. Starting from the Rabi model, we derive expressions for the TLS's oscillation frequencies and compare the results with those obtained from the model where the driving signal is treated classically. We show that in the limits of weak coupling and large photon number, the well-known expression for the Rabi frequency in the strong driving regime is recovered. In the opposite limit of strong coupling and small photon number, we find differences between the predictions of the semiclassical and quantum models. The results of the quantum picture can therefore be understood as Landau-Zener-Stueckelberg interferometry in the fully quantum regime.
\end{abstract}

%PACS numbers:

\maketitle

\section{Introduction}
\label{Sec:Introduction}

Landau-Zener-Stueckelberg (LZS) interferometry is encountered when a parameter of a quantum system are varied periodically in time such that the system repeatedly traverses an avoided crossing in its energy level diagram \cite{Shevchenko}. The response of the quantum system under such strong driving exhibits characteristic two-dimensional interference patterns that reflect the effects of interference involving the two traversals in a single driving period as well as the interference between operations corresponding to different driving periods. This situation has been the subject of numerous studies in recent years, covering both theory \cite{Garraway,Shytov,Creffield,Shevchenko2005,Ashhab2007,Son} and experiment \cite{Oliver,Berns,Sillanpaa,Saito2006,Wilson,Sun,VanDitzhuijzen,Petta,Childress,Zhou,Silveri,Neilinger}.

In the LZS problem, the driving field is generally treated classically, and only the driven two-level system is treated quantum mechanically. This picture can be considered a semiclassical approximation of a fully quantum treatment where the sinusoidal driving field is replaced by a harmonic oscillator that contains a large number of photons and hence behaves classically.

One can then ask the question: what happens when the driving field is treated quantum mechanically? In this paper we use the Rabi model to address this question. The closest that one can come to a classical signal in a quantum harmonic oscillator is a coherent state. Indeed we show that in the semiclassical limit with coherent states containing a large number of excitation quanta the predictions of the semiclassical and fully quantum models agree, although the results are described by seemingly different mathematical functions. One can expect that for coherent states with a small number of excitation quanta the discreteness of the oscillator's energy levels and the fluctuations in photon number relative to the average value start to have noticeable effects on the dynamics of the driven two-level system. We find that these are not the only deviations that the fully quantum model exhibits in relation to the semiclassical model. For example, starting with a well-defined photon number one could still obtain non-decaying sinusoidal oscillations but with a frequency that is completely different from what the semiclassical model predicts. We shall analyze this point and other similarities and differences in the predictions of the two models.

The remainder of this paper is organized as follows: in Sec.~\ref{Sec:SemiclassicalAndQuantumPictures} we introduce the semiclassical and fully quantum models for describing a driven two-level system. In Sec.~\ref{Sec:WeakDrivingCoupling} we review the case of weak driving. In Sec.~\ref{Sec:StrongDriving} we address the case of strong driving and LZS interferometry, as seen in the semiclassical and quantum models. We analyze the expressions for the Rabi frequency in the two models and compare the two expressions. In Sec.~\ref{Sec:TimeDomainSimulations} we present the results of time-domain simulations of the dynamics, giving a different perspective on the problem. We conclude with some final remarks in Sec.~\ref{Sec:Conclusion}.

\section{Semiclassical and fully quantum pictures}
\label{Sec:SemiclassicalAndQuantumPictures}

We consider a two-level quantum system  (to which we shall also refer as a qubit) driven by an external field with a sinusoidal time dependence. Specifically, we consider the Hamiltonian
\begin{equation}
\hat{H}_{\rm semiclassical} = -\frac{\Delta}{2} \hat{\sigma}_{x}-\frac{\epsilon+\hbar A\cos\left(\omega t+\phi_{\rm s}\right)}{2} \hat{\sigma}_{z},
\label{Eq:SemiclassicalHamiltonian}
\end{equation}
where $\Delta$ is the minimum energy gap at the avoided crossing point, $\epsilon$ is the average bias point relative to the so-called symmetry point ($\epsilon=0$), $A$, $\omega$ and $\phi_{\rm s}$ are, respectively, the amplitude, frequency and phase of the sinusoidal driving signal, and $\hat{\sigma}_{x,z}$ are qubit Pauli operators. When $\hbar A$ exceeds $\epsilon$ by an amount that is large compared to $\Delta$, the coefficient of the second term in the Hamiltonian oscillates between positive and negative values such that a sequence of Landau-Zener traversals is encountered, and the physics of LZS interferometry is realized.

The semiclassical Hamiltonian given in Eq.~(\ref{Eq:SemiclassicalHamiltonian}) can be seen as an approximation resulting from an underlying fully quantum Hamiltonian that treats both the driven two-level system and the driving field quantum mechanically. That underlying Hamiltonian is the Jaynes-Cummunigs (JC) Hamiltonian \cite{Jaynes}, which is also known as the Rabi-model Hamiltonian:
\begin{equation}
\hat{H}_{\rm quantum} = -\frac{\Delta}{2} \hat{\sigma}_x-\frac{\epsilon}{2} \hat{\sigma}_z+\hbar\omega \hat{a}^{\dagger}\hat{a}- \lambda \hat{\sigma}_z \left( \hat{a} + \hat{a}^{\dagger} \right),
\label{Eq:QuantumHamiltonian}
\end{equation}
where now $\omega$ is the frequency of a quantum harmonic oscillator (to which we shall also refer as the cavity whose excitations are photons) with annihilation and creation operators $\hat{a}$ and $\hat{a}^{\dagger}$, and $\lambda$ is the qubit-cavity coupling strength. It should be noted that the signs in this Hamiltonian were chosen such that they have a simple correspondence with those in Eq.~(\ref{Eq:SemiclassicalHamiltonian}) and that our results do not depend on this particular choice. Note also that we include the term proportional to $\hat{\sigma}_z$, which is sometimes omitted from the JC Hamiltonian.

When the cavity is treated classically, its field operators $\hat{a}$ and $\hat{a}^{\dagger}$ are replaced by the classical field values $\alpha e^{-i(\omega t + \phi_{\rm q})}$ and $\alpha e^{i(\omega t + \phi_{\rm q})}$, with $\alpha$ taken as a positive real number. Given a number of photons $n$ and noting that $n=\langle \hat{a}^{\dagger} \hat{a} \rangle$, the replacement of the quantum operators by the classical field values gives the amplitude of the classical field as $\alpha=\sqrt{n}$. As a result, in order to describe the same field intensity in the two models described by Eqs.~(\ref{Eq:SemiclassicalHamiltonian}) and (\ref{Eq:QuantumHamiltonian}), one must set
\begin{equation}
4\lambda\sqrt{n}=\hbar A.
\label{Eq:SemiclassicalQuantumFieldCorrespondence}
\end{equation}
In order to obtain fuller correspondence between the semiclassical and quantum treatments, the phases $\phi_{\rm s}$ and $\phi_{\rm q}$ must also be set to the same value, and we shall set both of them to zero for simplicity here.

\section{Resonant driving or coupling in the weak limit}
\label{Sec:WeakDrivingCoupling}

First let us consider the simple case of weak driving, which is the case studied by Rabi in Ref.~\cite{Rabi}. If we set $\hbar\omega=E_{\rm q}$ (where $E_{\rm q}=\sqrt{\epsilon^2+\Delta^2}$) and we assume that the driving is weak (i.e.~$\hbar A\ll E_{\rm q}$), we find that the two-level quantum system undergoes Rabi oscillations with frequency $(A/2)\times\cos\theta$, where $\theta=\tan^{-1}\left(\epsilon/\Delta\right)$.

Similarly we can take the JC Hamiltonian in the case $\hbar\omega=E_{\rm q}$, and with a simple rotation of qubit reference frame write it as
\begin{equation}
\hat{H}_{\rm quantum} = - \frac{E_{\rm q}}{2} \tilde{\sigma}_x+\hbar\omega \hat{a}^{\dagger}\hat{a} - \lambda \left( \cos\theta\tilde{\sigma}_z+\sin\theta\tilde{\sigma}_x \right) \left( \hat{a} + \hat{a}^{\dagger} \right).
\label{Eq:QuantumHamiltonianRotated}
\end{equation}
The weak-coupling regime in this model is given by the condition $\lambda\ll E_{\rm q}$ \cite{WeakCouplingFootnote}. In this section, as well as in parts of the following sections, we assume that the system is in the weak-coupling regime. In this regime the part of the coupling term in Eq.~(\ref{Eq:QuantumHamiltonianRotated}) that contains $\tilde{\sigma}_x$ does not affect the eigenvalues or eigenstates of the Hamiltonian to the lowest order that we need here, and it can be ignored. Ignoring it yields the Hamiltonian:
\begin{equation}
\hat{H}_{\rm quantum} = -\frac{E_{\rm q}}{2} \tilde{\sigma}_x+\hbar\omega \hat{a}^{\dagger}\hat{a} - \tilde{\lambda} \tilde{\sigma}_z \left( \hat{a} + \hat{a}^{\dagger} \right),
\end{equation}
where $\tilde{\lambda}=\lambda\cos\theta$. The angle $\theta$ here is defined exactly as in the semiclassical case described in the previous paragraph. For small $\tilde{\lambda}/E_{\rm q}$, we can further ignore the so-called counter-rotating terms and obtain the approximation
\begin{equation}
\hat{H}_{\rm quantum} = -\frac{E_{\rm q}}{2} \tilde{\sigma}_x+\hbar\omega \hat{a}^{\dagger}\hat{a} - \tilde{\lambda} \left( \tilde{\sigma}_+ \hat{a} + \tilde{\sigma}_- \hat{a}^{\dagger} \right),
\label{Eq:QuantumHamiltonianRWA}
\end{equation}
where $\tilde{\sigma}_{\pm}=\left( \tilde{\sigma}_z \pm i \tilde{\sigma}_y \right)/2$ and the effect of these operators is to excite or de-excite the qubit: $\tilde{\sigma}_+ \ket{g} = \ket{e}$ and $\tilde{\sigma}_- \ket{e} = \ket{g}$ where $\ket{g}$ and $\ket{e}$ are the ground and excited states of the the bare qubit Hamiltonian (i.e.~$\hat{H}_{\rm q}=-E_{\rm q}\tilde{\sigma}_x/2$).

The ground state of the Hamiltonian in Eq.~(\ref{Eq:QuantumHamiltonianRWA}) is given by $\ket{g,0}$, where the first and second indices specify, respectively, the state of the qubit and the number of excitations in the oscillator. Apart from the ground state, the low energy levels are grouped into pairs with energies $n\hbar\omega\pm \tilde{\lambda}\sqrt{n+1}$ above the ground state energy, with corresponding eigenstates given by $\left( \ket{g,n+1}\mp\ket{e,n}\right)/\sqrt{2}$. The fact that the energy eigenstates are quantum superpositions of this form naturally leads to Rabi-like oscillations. If, for example, the system is initially set in a state with the qubit in its ground state $g$ and the cavity in a state with $n$ photons, and hence the combined system is initially in the state $\ket{g,n}$, the system will undergo oscillations between the states $\ket{g,n}$ and $\ket{e,n-1}$ with frequency $2\tilde{\lambda}\sqrt{n}$.

Obviously the two descriptions above, with oscillation frequencies given by $A/2\times\cos\theta$ and $2\lambda\sqrt{n}\cos\theta$, are essentially equivalent and give the same value for the Rabi frequency when we set $4\lambda\sqrt{n}=\hbar A$.

\section{Strong driving}
\label{Sec:StrongDriving}

We now turn to the case of strong driving. We start with the semiclassical model. When we set $k\hbar\omega=\epsilon$ with $\omega\gg\Delta$, we obtain Rabi oscillations with frequency
\begin{equation}
\Omega_{\rm Rabi,s} = \frac{\Delta}{\hbar} J_k\left(\frac{A}{\omega}\right),
\label{Eq:RabiFrequencySemiclassical}
\end{equation} 
where $J_k$ is the $k$-th order Bessel function of the first kind \cite{Shirley}. This behaviour was observed recently using superconducting qubits \cite{Nakamura,Saito2004}. Note that although Eq.~(\ref{Eq:RabiFrequencySemiclassical}) does not contain $\epsilon$ explicitly, this parameter is of course crucial in determining that the driving is resonant and therefore in determining the value of $k$. Note also that here we consider only the case where the resonance condition $k\hbar\omega=\epsilon$ is satisfied, because this is the case when the qubit exhibits the characteristic LZS-Rabi oscillation dynamics \cite{Shevchenko}.

If we want to investigate the same situation in the fully quantum picture, we can start by writing the Hamiltonian in the form
\begin{equation}
\hat{H}_{\rm quantum} = \hat{H}_0 + \hat{H}_1,
\end{equation}
where
\begin{eqnarray}
\hat{H}_0 & = & -\frac{\epsilon}{2} \hat{\sigma}_z+\hbar\omega \hat{a}^{\dagger}\hat{a} - \lambda \hat{\sigma}_z \left( \hat{a} + \hat{a}^{\dagger} \right) \nonumber \\
\hat{H}_1 & = & -\frac{\Delta}{2} \hat{\sigma}_x,
\end{eqnarray}
which leads to the generalized rotating-wave approximation described in Ref.~\cite{Irish}. The eigenstates of $\hat{H}_0$ are given by
\begin{eqnarray}
\ket{\uparrow}\otimes\hat{D}\left(\frac{\lambda}{\hbar\omega}\right)\ket{n} \;\;\; {\rm and} \;\;\;
\ket{\downarrow}\otimes\hat{D}\left(-\frac{\lambda}{\hbar\omega}\right)\ket{n},
\label{Eq:GRWABasisStates}
\end{eqnarray}
with respective energies $\mp\frac{\epsilon}{2}+n\hbar\omega-\frac{\lambda^2}{\hbar\omega}$. Here the displacement operator $\hat{D}(x)=e^{x\left(\hat{a} - \hat{a}^{\dagger}\right)}$, and we have used the state definitions $\hat{\sigma}_z\ket{\uparrow}=\ket{\uparrow}$ and $\hat{\sigma}_z\ket{\downarrow}=-\ket{\downarrow}$. When $\epsilon=k\hbar\omega$, the Hamiltonian $\hat{H}_0$ has degeneracies between the states $\ket{\uparrow}\otimes\hat{D}\left(\frac{\lambda}{\hbar\omega}\right)\ket{n+k}$ and $\ket{\downarrow}\otimes\hat{D}\left(-\frac{\lambda}{\hbar\omega}\right)\ket{n}$. These degeneracies are lifted by $\hat{H}_1$ with the splittings, and hence Rabi oscillation frequencies, given by
\begin{equation}
\Omega_{\rm Rabi,q} = \frac{\Delta}{\hbar} e^{-2\lambda^2/(\hbar\omega)^2} \left(\frac{2\lambda}{\hbar\omega}\right)^k \sqrt{\frac{n!}{(n+k)!}} L_{n}^{k}\left[\left(\frac{2\lambda}{\hbar\omega}\right)^2\right],
\label{Eq:RabiFrequencyQuantum}
\end{equation}
where $L_{n}^k$ are associated Laguerre polynomials.

\begin{figure}[h]
\includegraphics[width=7.5cm]{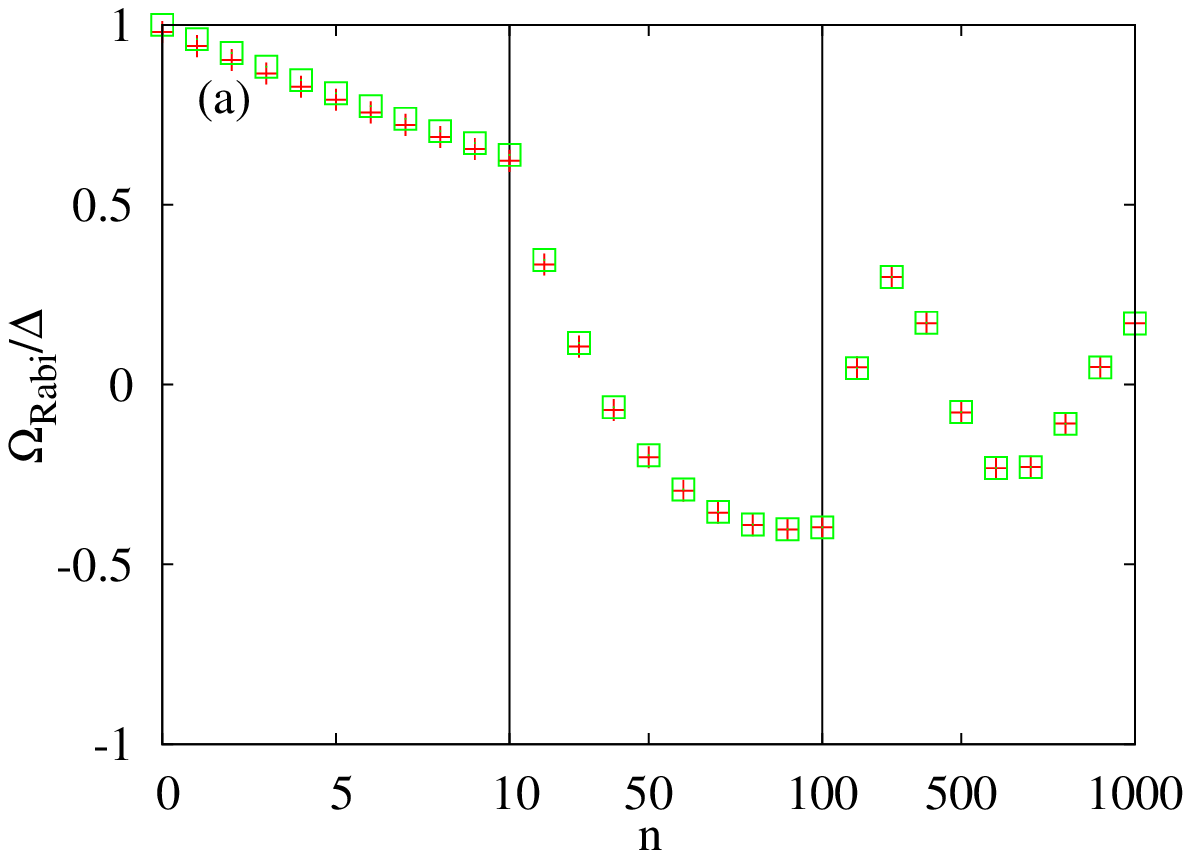}
\includegraphics[width=7.5cm]{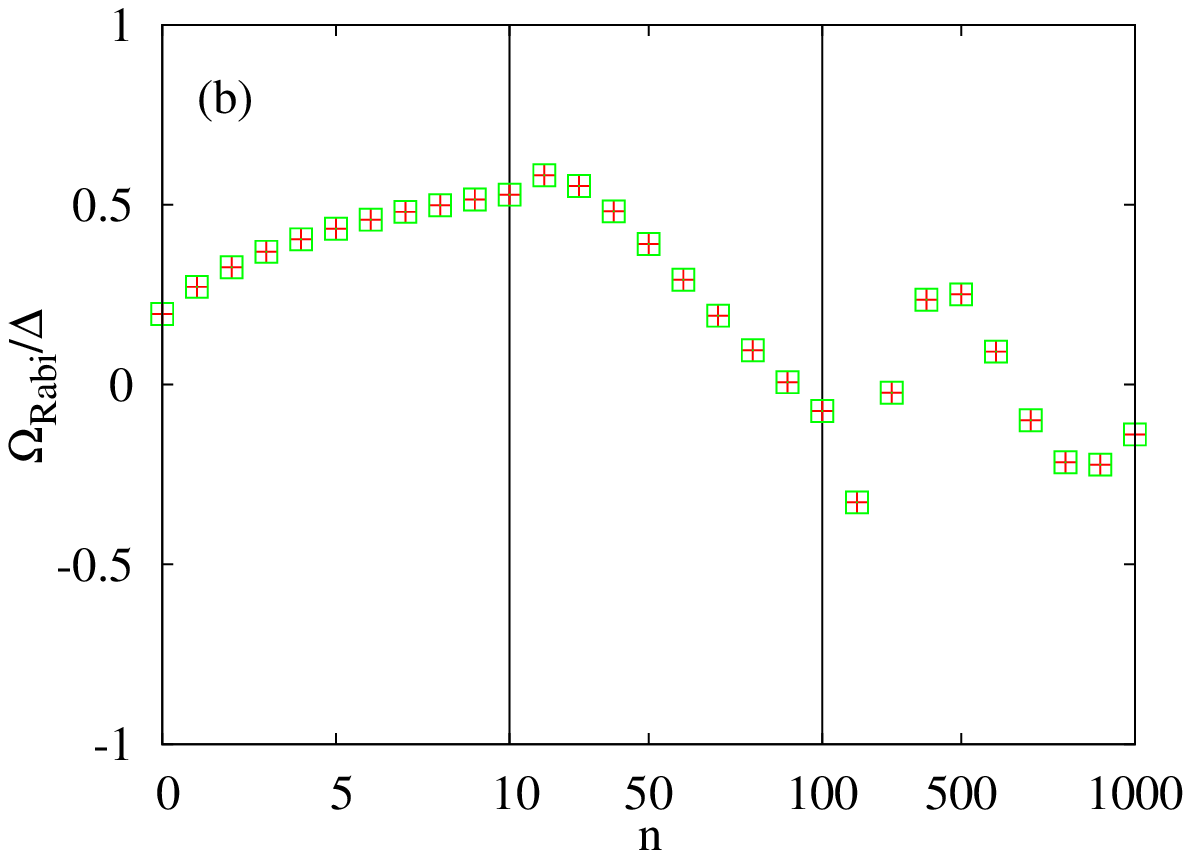}
\includegraphics[width=7.5cm]{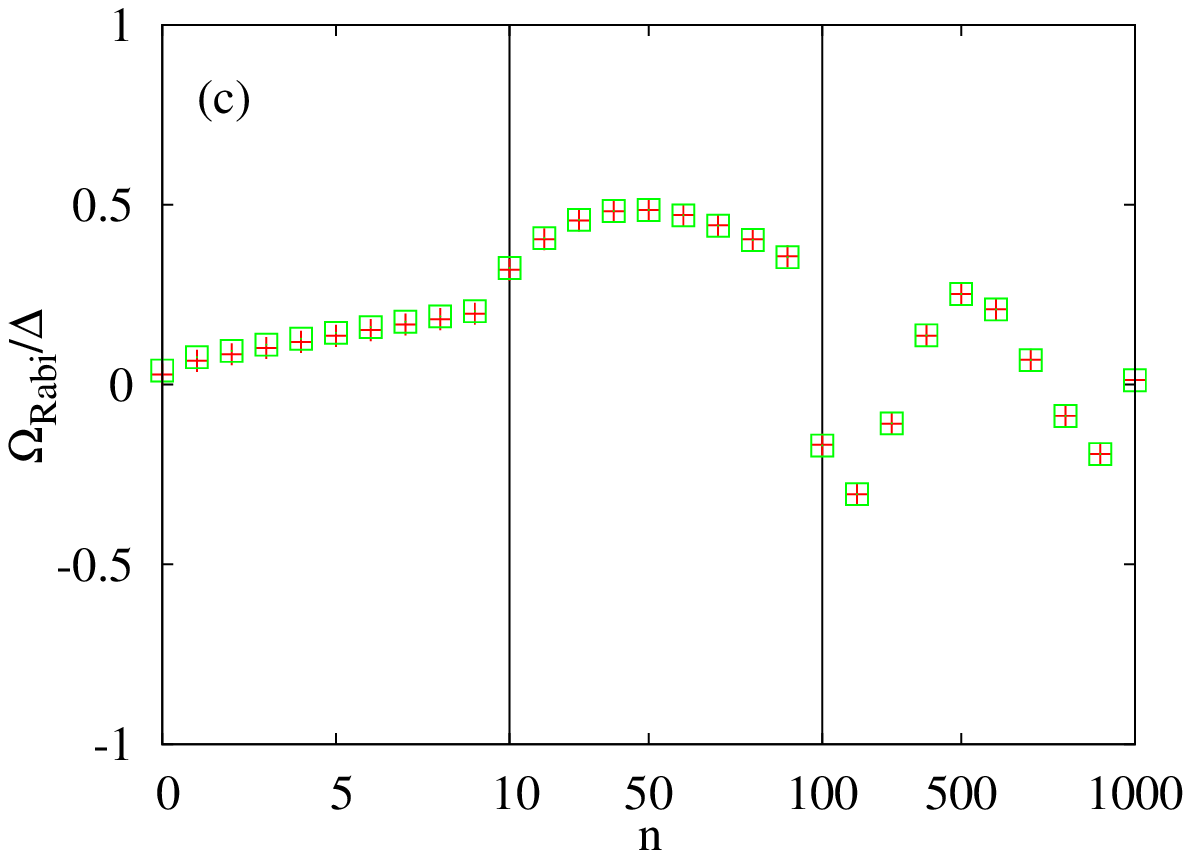}
\includegraphics[width=7.5cm]{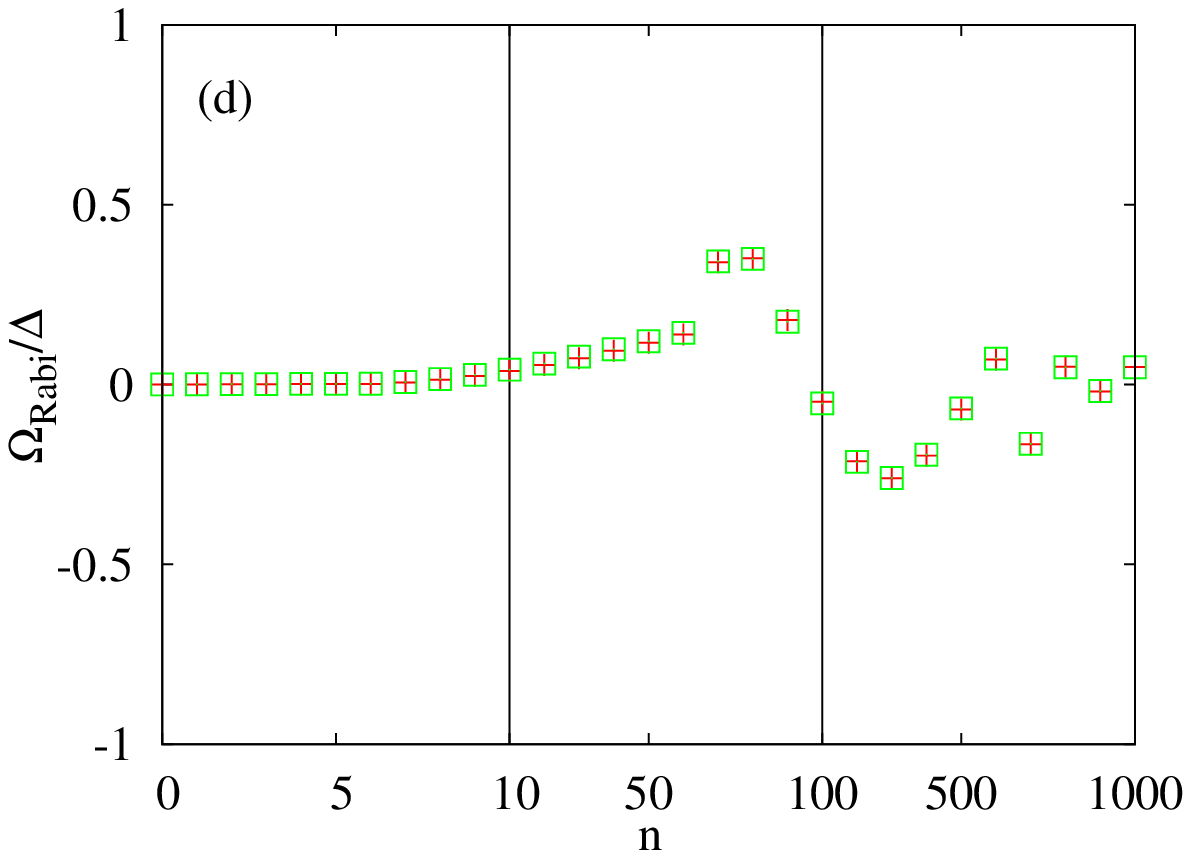}
\caption{Rabi frequencies obtained from the semiclassical model (squares, calculated using Eq.~\ref{Eq:RabiFrequencySemiclassical}) and from the fully quantum model (+ signs, calculated using Eq.~\ref{Eq:RabiFrequencyQuantum}) as functions of photon number $n$. In all the panels we set $\lambda/(\hbar\omega)=0.1$, which corresponds to the weak-coupling regime. The different panels correspond to different values of $k$ in the resonance condition $k\hbar\omega=\epsilon$. In particular we take the cases $k=0$ (a), 1 (b), 2 (c) and 5 (d). The x axis is divided into three parts that have different scales: the first part ranging from 0 to 10, then second from 10 to 100 and the third from 100 to 1000. We note here that the frequency should be obtained by taking the absolute value of the relevant expressions, but we keep the signs here in order to capture what can be considered accidental agreement between the two calculations in case they give the same value with opposite signs.}
\label{Fig:RabiFrequencyWeakCoupling}
\end{figure}

We can now compare the predictions of the semiclassical and quantum models by comparing the two expressions given in Eqs.~(\ref{Eq:RabiFrequencySemiclassical}) and (\ref{Eq:RabiFrequencyQuantum}). An important point to note here is that, because of the relation in Eq.~(\ref{Eq:SemiclassicalQuantumFieldCorrespondence}), there are two ways to obtain a large effective driving field in the quantum model, namely by having a large value of either $\lambda$ or $n$ such that their product is comparable to or larger than $\hbar\omega$.

We start with the case of weak coupling between the qubit and the cavity (i.e.~$\lambda/(\hbar\omega)\ll 1$), where strong driving would require a large value of $n$. In Fig.~\ref{Fig:RabiFrequencyWeakCoupling} we plot the Rabi frequency as obtained from the semiclassical and fully quantum calculations for $\lambda/(\hbar\omega)=0.1$ for the four cases $k=0,1,2$ and 5. In fact in some recent studies, e.g.~Refs.~\cite{Niemczyk,FornDiaz2010}, the value $\lambda/(\hbar\omega)=0.1$ has been identified as being in the ultrastrong-coupling regime of the Rabi model. However, for purposes of this study this value of $\lambda/(\hbar\omega)$ can be considered to lie in the weak-coupling regime because it leads to the same behaviour as what we would obtain for very small values of $\lambda/(\hbar\omega)$. We can see from Fig.~\ref{Fig:RabiFrequencyWeakCoupling} that there is excellent agreement between the semiclassical and quantum models in their prediction of the Rabi frequency for all values of $n$, at least up to $k=5$.

\begin{figure}[h]
\includegraphics[width=7.5cm]{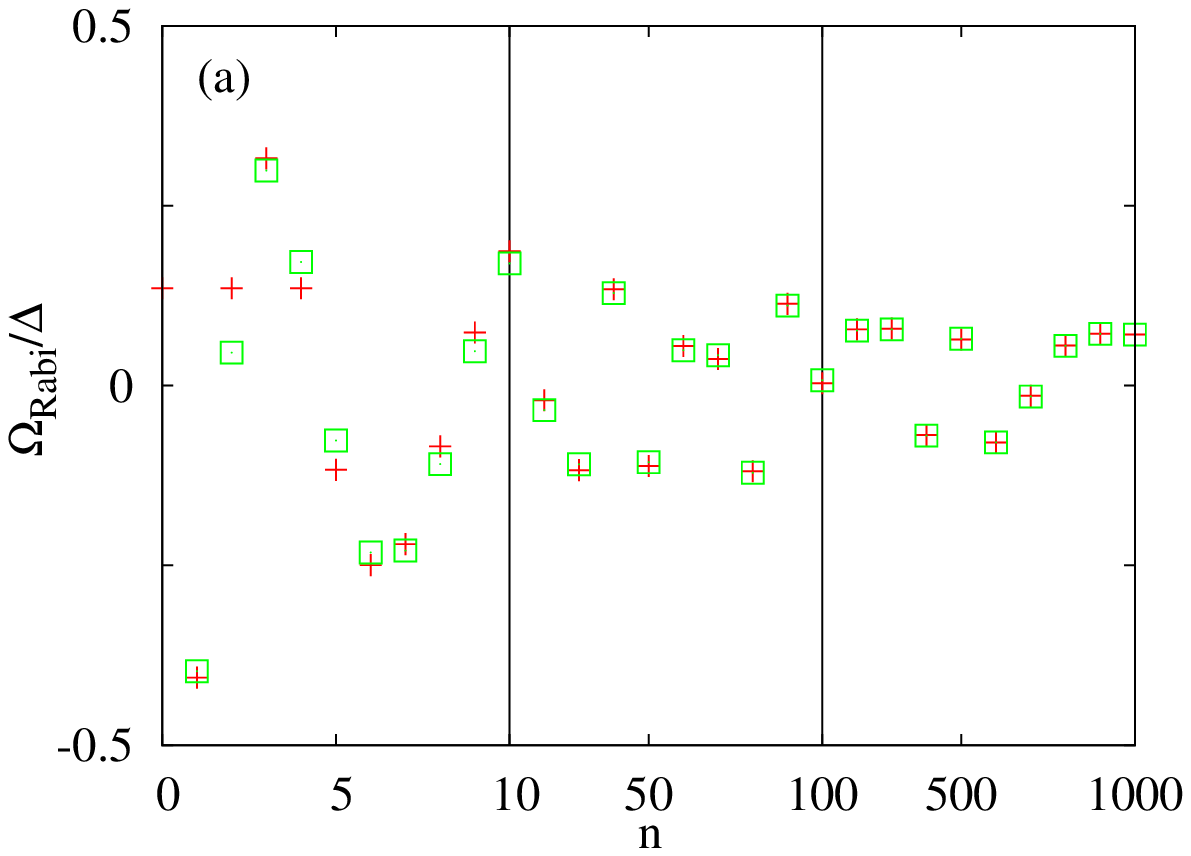}
\includegraphics[width=7.5cm]{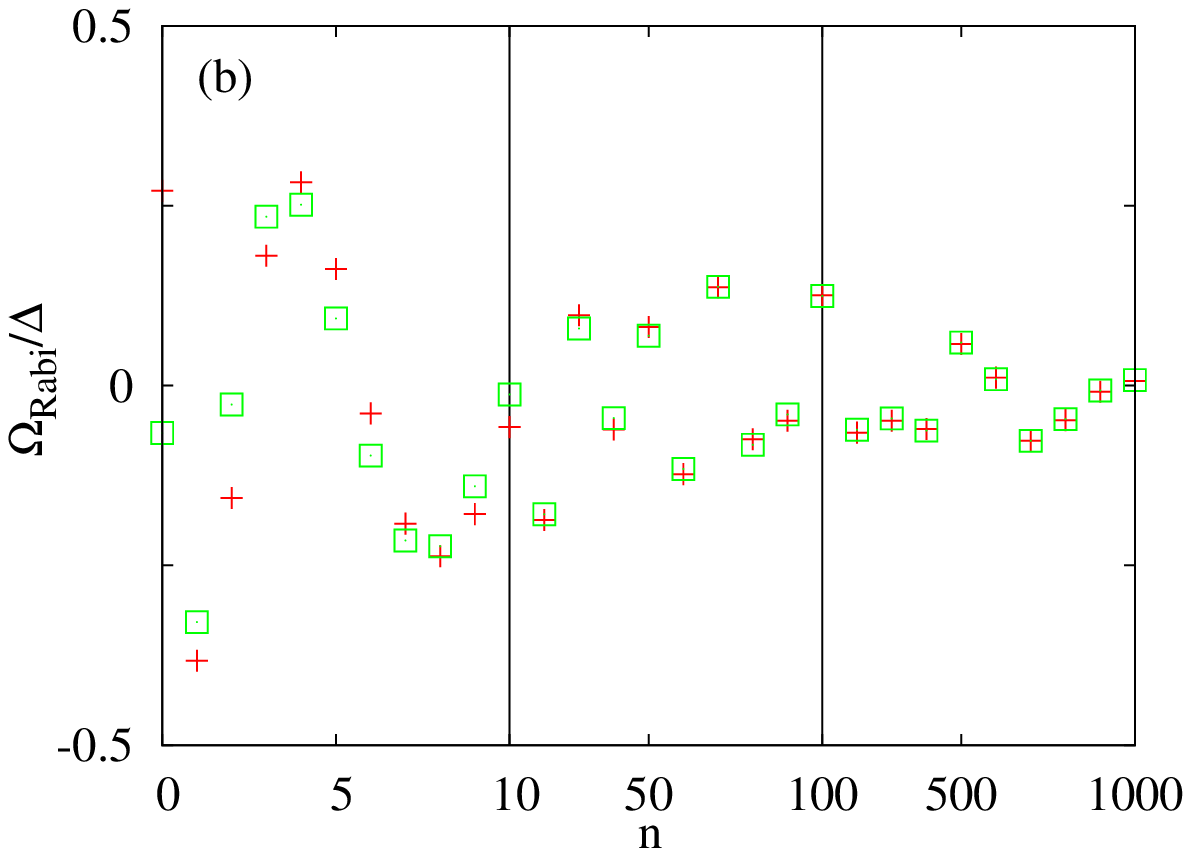}
\includegraphics[width=7.5cm]{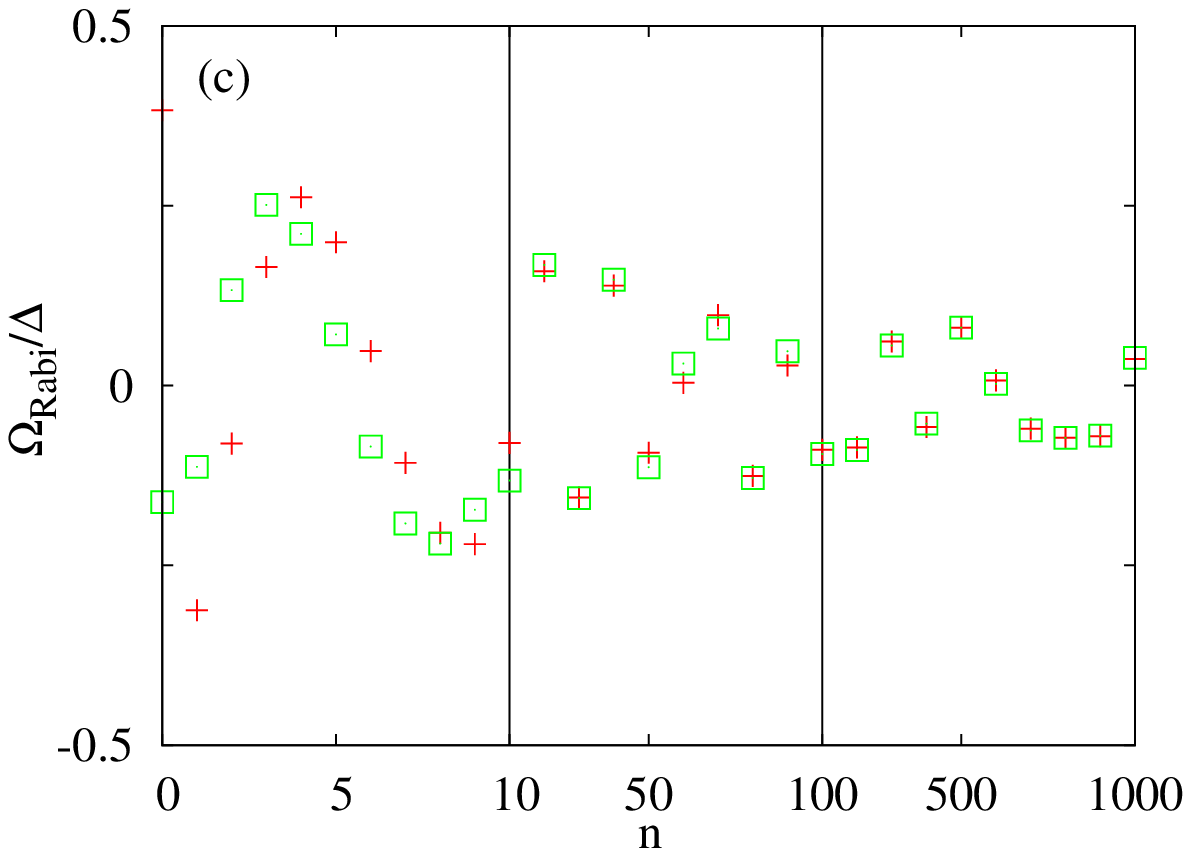}
\includegraphics[width=7.5cm]{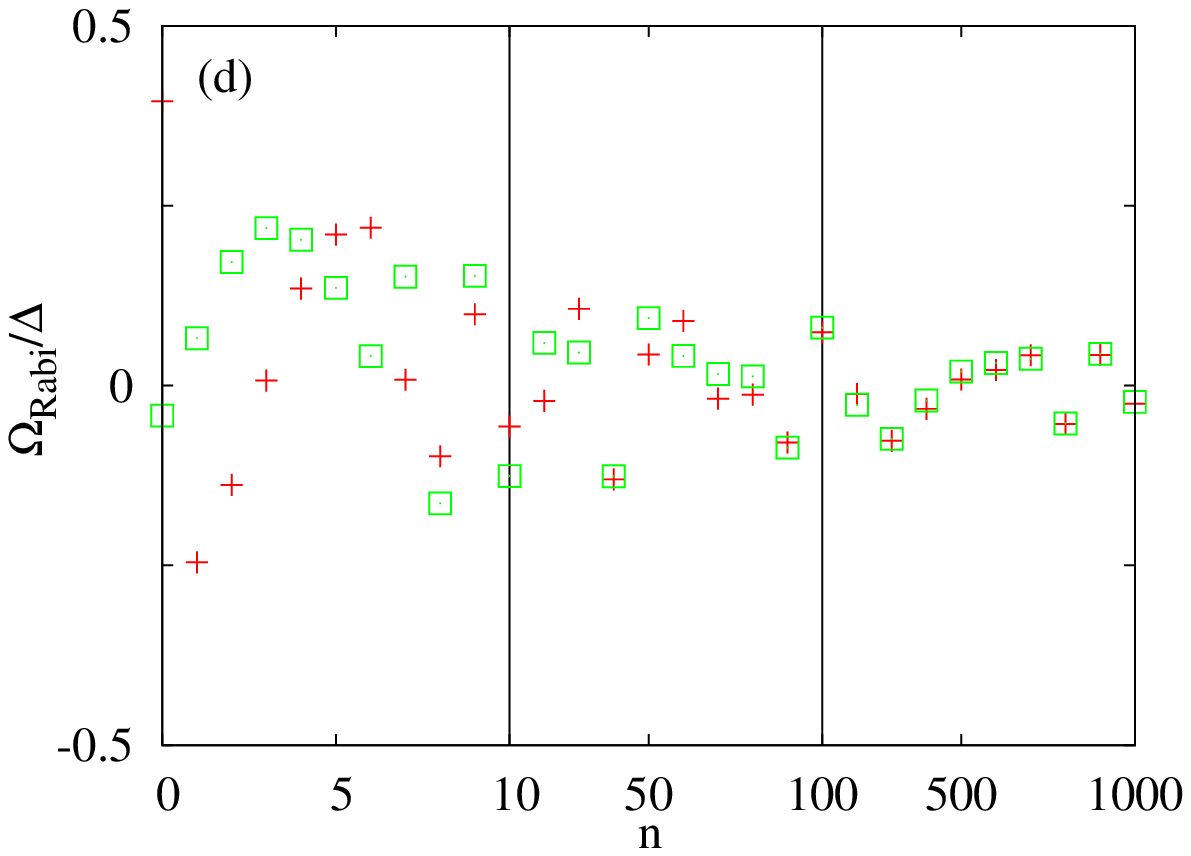}
\caption{Same as in Fig.~\ref{Fig:RabiFrequencyWeakCoupling}, but with $\lambda/(\hbar\omega)=1$, which corresponds to the strong-coupling regime.}
\label{Fig:RabiFrequencyIntermediateCoupling}
\end{figure}

Next we consider the case of strong coupling between the cavity and the qubit, and we set $\lambda/(\hbar\omega)=1$. This value corresponds to the so-called deep-strong-coupling regime of the Rabi model \cite{Casanova,Wolf,Yoshihara,YoshiharaSpectra,Ashhab2010}. The two expressions for the Rabi frequency are plotted in Fig.~\ref{Fig:RabiFrequencyIntermediateCoupling}. The two expressions agree at large values of $n$, but we now see clear deviations at small values of $n$. The deviations also extend to larger values of $n$ with increasing values of $k$. For example, for $k=0$ the two expressions start to agree very well when $n\gtrsim 50$, whereas we need $n\gtrsim 500$ for $k=5$. Furthermore, below $n=10$ it seems that the two calculations sometimes give completely different results.

\begin{figure}[h]
\includegraphics[width=7.5cm]{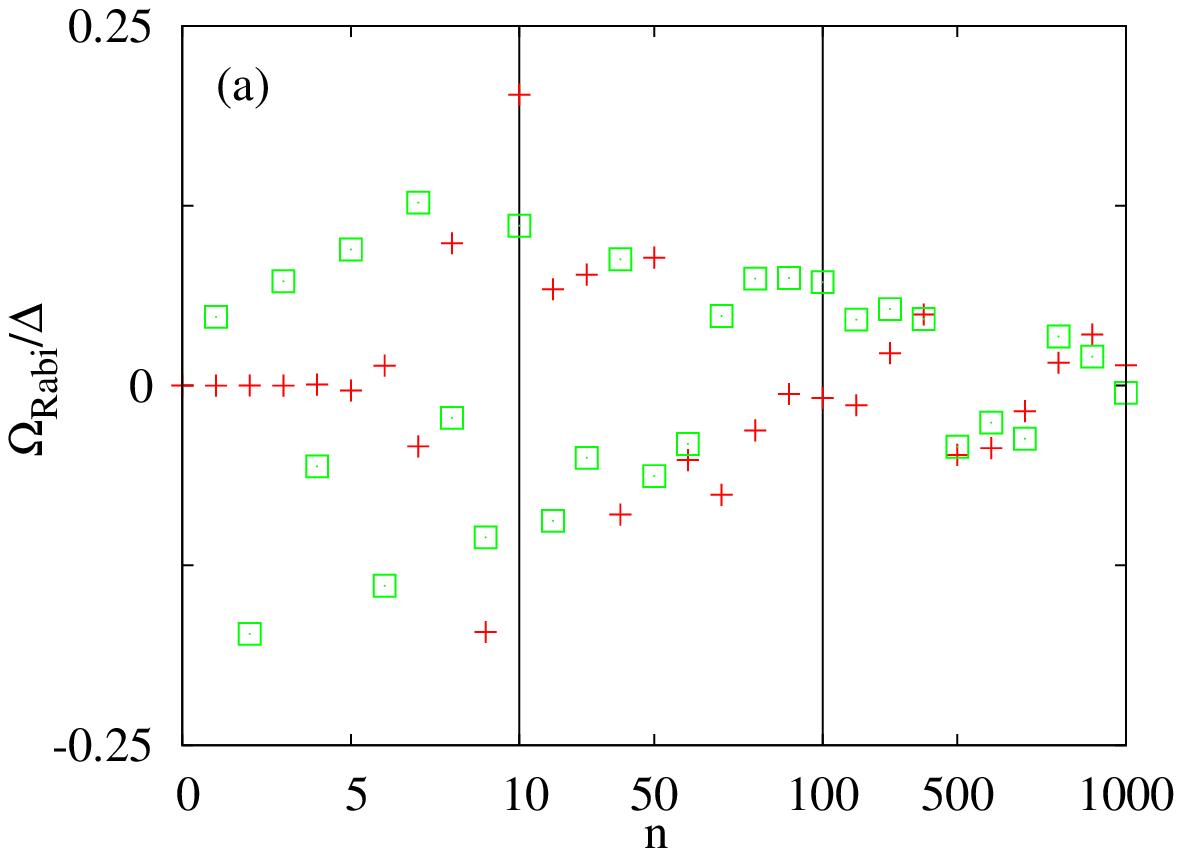}
\includegraphics[width=7.5cm]{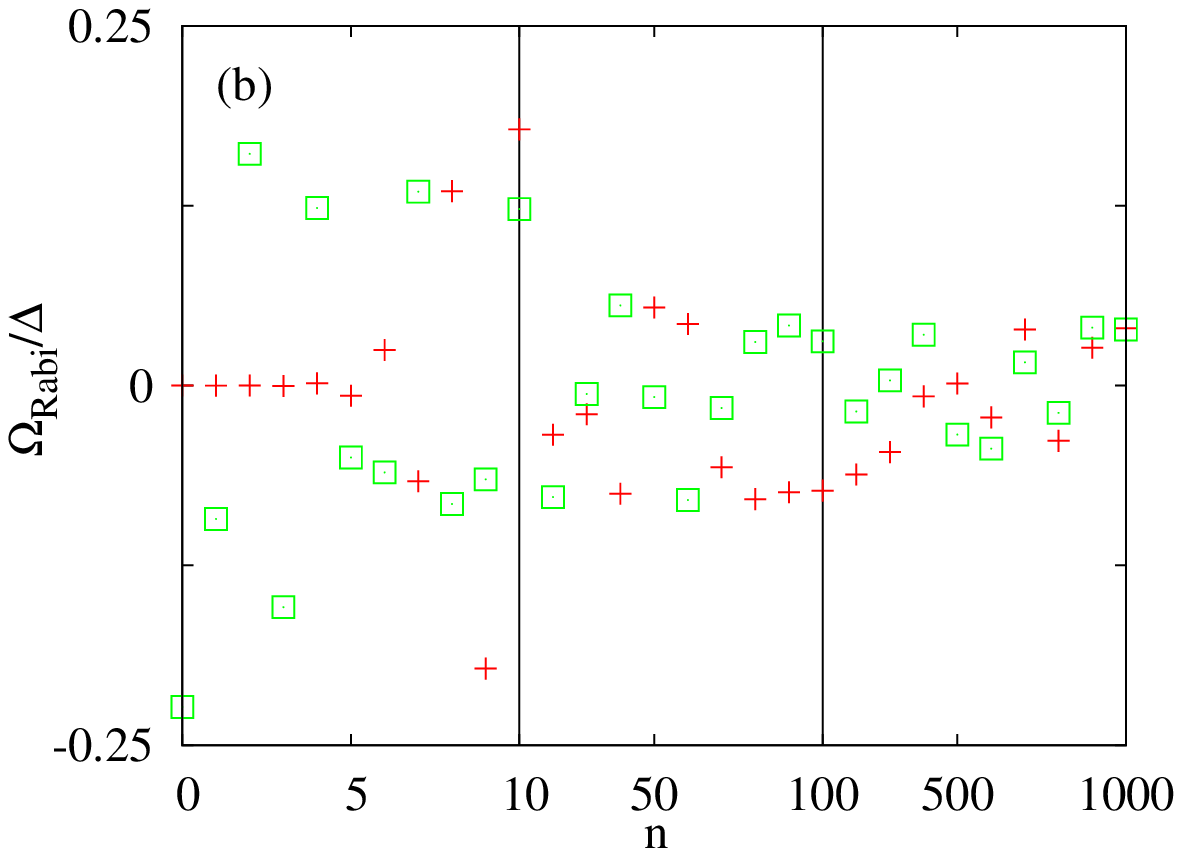}
\includegraphics[width=7.5cm]{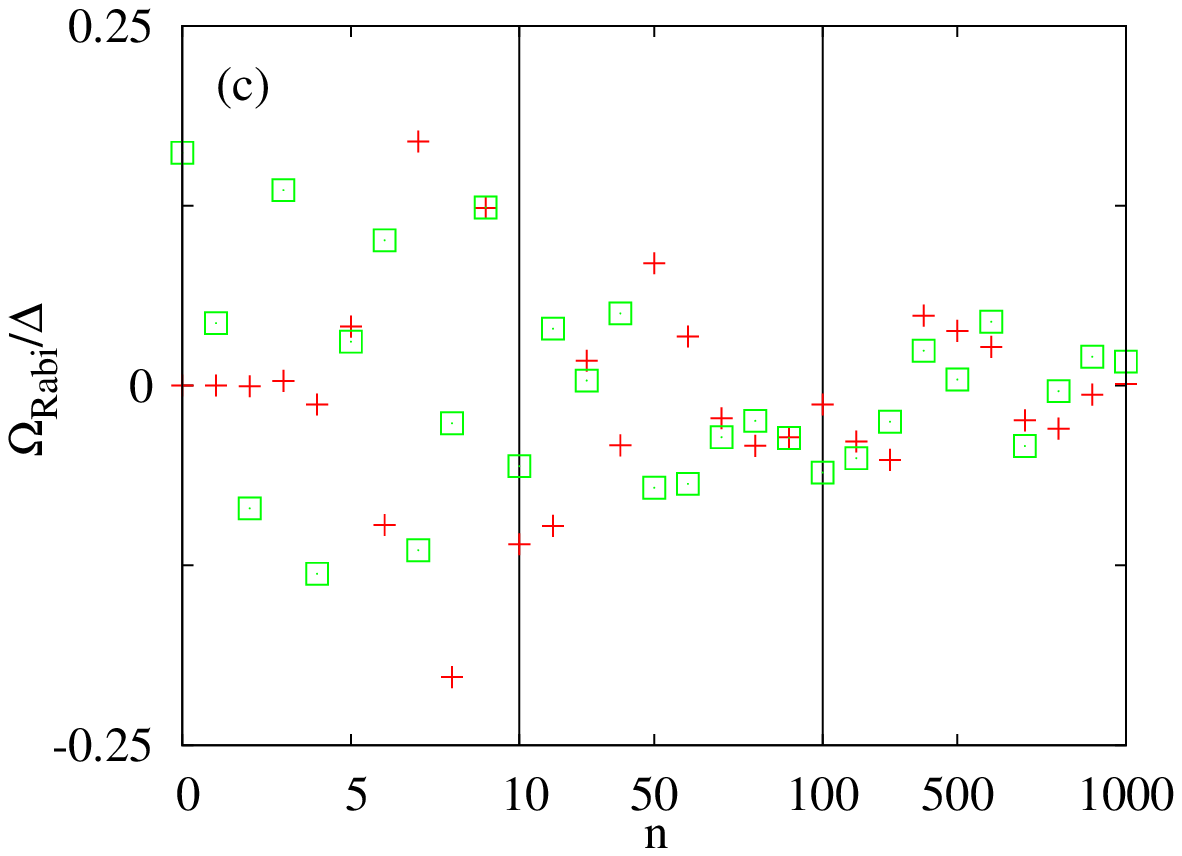}
\includegraphics[width=7.5cm]{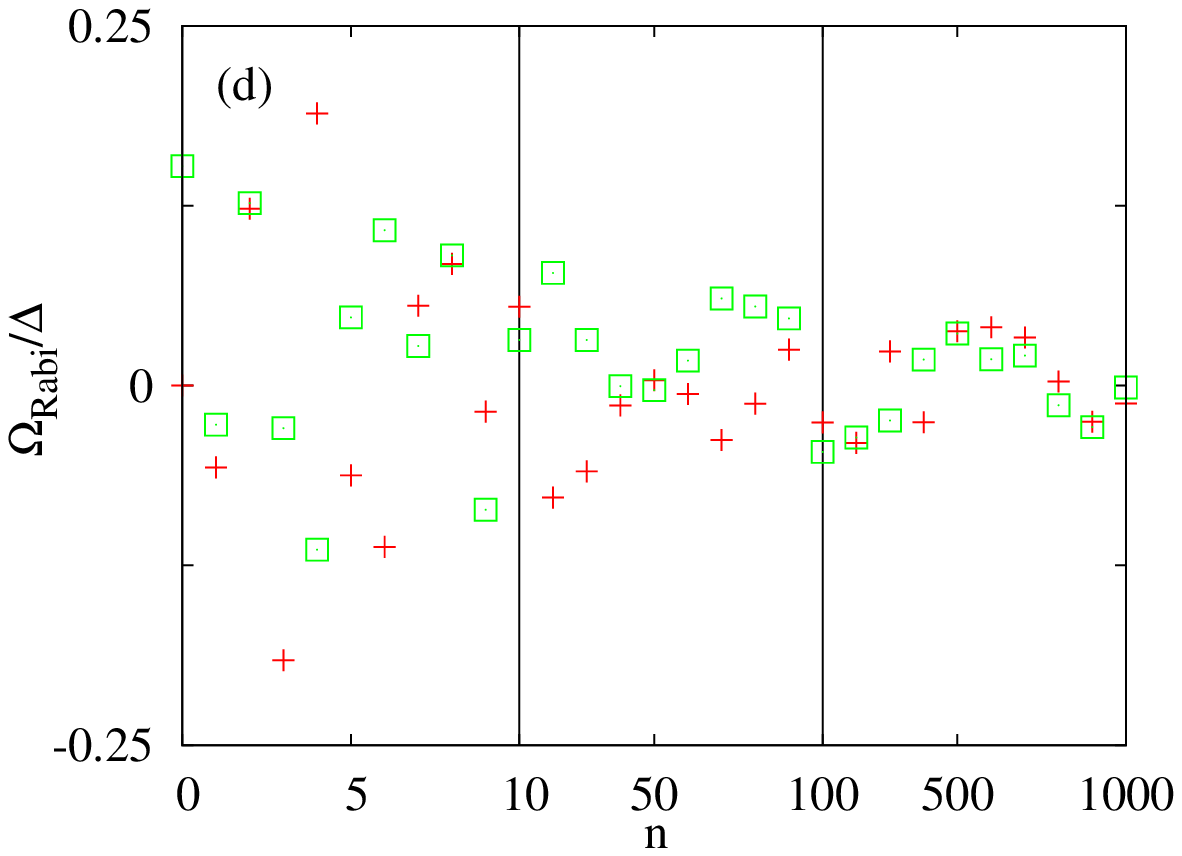}
\caption{Same as in Fig.~\ref{Fig:RabiFrequencyWeakCoupling}, but with $\lambda/(\hbar\omega)=3$, which corresponds to extremely strong coupling.}
\label{Fig:RabiFrequencyStrongCoupling}
\end{figure}

If we push the qubit-cavity coupling strength to even larger values, specifically $\lambda/(\hbar\omega)=3$ in Fig.~\ref{Fig:RabiFrequencyStrongCoupling}, then even for $k=0$ the two expressions for the Rabi frequency start to agree only when $n\gtrsim 10^4$ (not shown in the figure). One interesting feature that we can see in Fig.~\ref{Fig:RabiFrequencyStrongCoupling} is that for $n\leq 4$ and small $k$ the expression derived from the quantum model is essentially zero, while the semiclassical derivation gives a finite value for the Rabi frequency.

A conclusion that we can draw from Figs.~\ref{Fig:RabiFrequencyWeakCoupling}-\ref{Fig:RabiFrequencyStrongCoupling} is that the semiclassical and quantum models give the same results for the Rabi frequency in the limits of small $\lambda/(\hbar\omega)$ and large $n$. Based on this observation, we can use Eqs.~(\ref{Eq:RabiFrequencySemiclassical}) and (\ref{Eq:RabiFrequencyQuantum}) to deduce the approximation
%
%\begin{widetext}
\begin{equation}
J_k\left(4x\sqrt{n}\right) \approx e^{-2x^2} \left(2x\right)^k \sqrt{\frac{n!}{(n+k)!}} L_{n}^{k}\left[4x^2\right],
\end{equation}
%\end{widetext}
%
which is valid for small $x$ and/or large $n$. It is interesting that considering the same physical problem from two different perspectives has led us to infer a relationship between two mathematical functions that are not obviously related. A similar situation is also given in the appendix.

It is also worth noting here that recently an expression containing two Bessel functions was derived for the Rabi frequency in the case of $\epsilon=0$ and $\hbar\omega=\Delta$ \cite{Deng,Lu}. It is not obvious how this expression can be derived from the quantum model.

When considering the dynamics in the fully quantum picture, it is also interesting to consider the back-action of the qubit on the driving field. In the semiclassical picture, the field is an externally given function of time that is not affected by the state of the driven system. In the quantum picture, any change in the state of the qubit will be accompanied by a change in the state of the cavity. For example, the excitation of the qubit from the ground to the excited state in a $k$-photon resonance will be accompanied by the absorption of $k$ photons from the cavity. The alert reader might have already noticed that the expression that we used for the semiclassical model contains the photon number $n$, even though the oscillations involve alternation between $n$ and $n+k$ photons. Taking this point into consideration, one might think that it would be better to set the classical field $\hbar A$ to $4\lambda\sqrt{n+k/2}$ instead of $4\lambda\sqrt{n}$. A closer inspection of the functions plotted in Figs.~\ref{Fig:RabiFrequencyWeakCoupling}-\ref{Fig:RabiFrequencyStrongCoupling} reveals that the situation is somewhat more complicated. For $\lambda/(\hbar\omega)=0.1$, taking large values of $n$ we find that we obtain the closest agreement between the semiclassical and quantum calculations by setting $\hbar A=4\lambda\sqrt{n+k/2+0.5}$ for all four values of $k$ plotted in Fig.~\ref{Fig:RabiFrequencyWeakCoupling}. The term $k/2$ is therefore consistently there. However, we also have the additional 0.5 whose origin is not clear. From fitting the data for all the combinations of $\lambda/(\hbar\omega)\in\{0.1,1,2,3\}$ and $k\in\{0,1,2,5\}$ we consistently find that the best agreement is obtained when we set $\hbar A=4\lambda\sqrt{n+k/2+0.5-[\lambda/(\hbar\omega)]^2/3}$. We do not know the origin  of the last two terms inside the square-root. We also note that for small values of $n$ this formula gave good agreement for small values of $\lambda/(\hbar\omega)$, but deviations between the two expressions persisted, especially for $\lambda/(\hbar\omega)\geq 2$, where no value of the shift seemed to consistently reduce the deviations.

Another back-action effect arises naturally if for a moment we consider what happens when we choose parameters that do not satisfy resonance conditions. In this case the energy eigenstates are divided into two groups corresponding to the qubit states $\ket{\uparrow}$ and $\ket{\downarrow}$, as described by Eq.~(\ref{Eq:GRWABasisStates}). The cavity part of each one of these energy eigenstate is described by a Fock state, just as in an isolated harmonic oscillator, but with a qubit-state-dependent displacement. If the qubit is prepared in one of the states $\ket{\uparrow}$ and $\ket{\downarrow}$ and the cavity is initially prepared in a coherent state, then the cavity field will undergo oscillations with frequency $\omega$ about a qubit-state-dependent equilibrium point that is shifted from the origin by $\pm\lambda/(\hbar\omega)$. Even if we choose parameters that satisfy a resonance condition, these oscillations will occur on short timescales. On longer timescales, specifically the timescale of Rabi oscillations, the qubit will undergo oscillations between the states $\ket{\uparrow}$ and $\ket{\downarrow}$, and the cavity field will adjust its oscillation pattern such that it remains correlated with the qubit's state. For example, after a full transfer of probability from the state $\ket{\uparrow}$ to the state $\ket{\downarrow}$ or vice versa the cavity field will have shifted the origin of its oscillations by $2\lambda/(\hbar\omega)$. As discussed above, the amplitude of the cavity field oscillations will also change because of the absorption or emission of $k$ photons in a $k$-photon resonance.

\section{Time-domain simulations}
\label{Sec:TimeDomainSimulations}

We now present the results of simulations where we solve the Schr\"odinger equation and obtain the qubit state populations as functions of time in the semiclassical and fully quantum models. We take the case of two-photon resonance $\epsilon=2\hbar\omega$. We choose a driving amplitude $A/\omega=10$ (or an equivalent value of $n$ in the quantum model), which gives constructive interference and therefore a reasonably high Rabi frequency. For the qubit gap we choose the value $\Delta/(\hbar\omega)=0.4$, which is relatively large and clearly shows the step-like dynamics that is one of the characteristics of LZS interferometry \cite{Zhou}.

\begin{figure}[h]
\includegraphics[width=8.0cm]{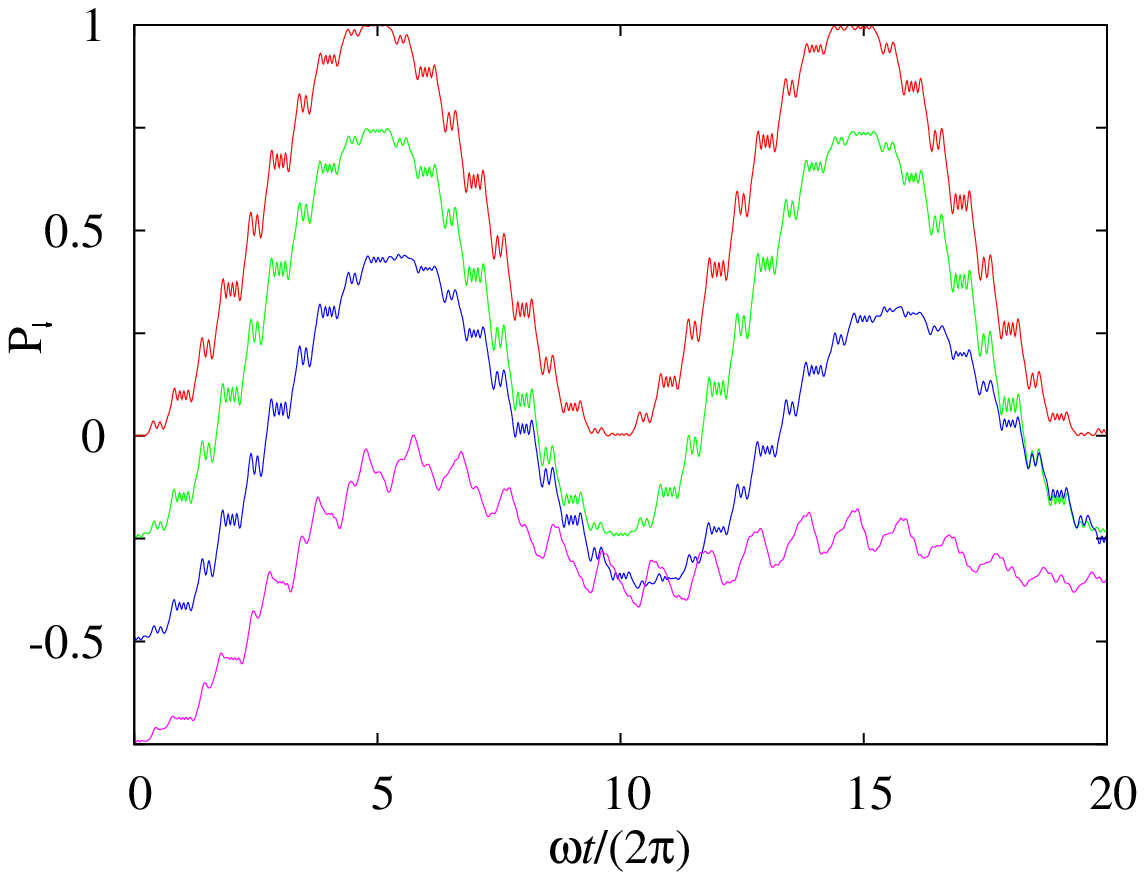}
\caption{Occupation probability $P_{\downarrow}$ of the qubit state $\ket{\downarrow}$ as a function of time $t$ with driving amplitude $A/\omega=10$, qubit bias $\epsilon/(\hbar\omega)=2$ and qubit gap $\Delta/(\hbar\omega)=0.4$. The red line is obtained from the semiclassical model. The green line, which is shifted down by 0.25 in order to make it resolvable from the red line, is obtained from the quantum model with the cavity initially set in a coherent state with $\langle n \rangle=1000$, and the coupling strength $\lambda$ is chosen such that $n=1000$ corresponds to $A/\omega=10$. There red and green line agree very well and would hardly be resolvable without the shift. The blue line, which is shifted down by 0.5, is obtained from the quantum model with $\langle n \rangle=100$. It shows a clear deviation from the other two lines. In particular, it exhibits a decay that can be attributed to the significant spread in the Rabi frequency for the different $n$ values in the coherent state. The magenta line, which is shifted down by 0.75, is obtained from the quantum model with $\langle n \rangle=10$. Instead of the step-like dynamics, we can now see that there are only a few frequency components in the dynamics, as would be expected for such a small value of $\langle n \rangle$.}
\label{Fig:TimeDomainDynamics}
\end{figure}

The results of the simulations are plotted in Fig.~\ref{Fig:TimeDomainDynamics}. When we use the quantum model with a coherent state that contains $\sim$ 1000 photons, the results of the quantum model are almost indistinguishable from those of the semiclassical model. When we extend the simulations to longer timescales, we see a decay in the oscillations with a decay time of $\sim 30\times 2\pi/\Omega_{\rm Rabi}$. This value for the decay time is rather large considering that the quantum fluctuations in $n$ in such a coherent state are $\sqrt{1000}$, which is about 3\% of the mean photon number. The reason for this weak decay is that the point $A/\omega=10$ is very close to a maximum in $J_2(A/\omega)$ and fluctuations up to 3\% in $A$ do not result in large fluctuations in $\Omega_{\rm Rabi}$.

When we change the parameters such that we now have $\sim$ 100 photons in the coherent state, we see very fast decay in the oscillations, with a decay time of only a few times the Rabi oscillation period. The quantum fluctuations in $n$ are now on the order of 10\%, which means that the Rabi frequency varies significantly for the different Fock states that make up this coherent state. Interestingly, if we use a Fock state with exactly 100 photons and the coupling strength adjusted to the appropriate value, we recover decay-less oscillations with a frequency that is essentially identical to that obtained from the semiclassical calculation. This result is somewhat surprising, because coherent states are generally expected to give the closest similarity to classical fields.

When we reduce the average number of photons in the coherent state to 10, the decay becomes so fast that we can barely see the Rabi oscillations, and instead of the step-like dynamics that is characteristic of LZS interferometry we now see only a few frequency components in the dynamics.

\section{Conclusion}
\label{Sec:Conclusion}

We have treated the problem of Landau-Zener-Stueckelberg interferometry when the driving field is in the quantum regime. We have found that the quantum treatment reproduces the results of the semiclassical treatment when the qubit-cavity coupling strength is small or the number of photons in the cavity is large. In this case the expressions containing Laguerre polynomials that are typical in the study of the Rabi model coincide with the expressions that contain Bessel functions and are typical in strong-driving problems. The semiclassical expression is no longer applicable, however, for large values of $\lambda/(\hbar\omega)$ and small values of $n$. In this case, not only do quantum fluctuations cause decaying oscillations in the time-domain, but the $n$ dependence of the Rabi frequency itself shows differences in the predictions of the two models.

In our simulations of the Rabi model we have relied on numerical diagonalization of the Hamiltonian. It should be noted that the recent advances made on the integrability and solution of the Rabi model provide additional analytical tools and allow the simulation of very large system using reasonable computational resources \cite{Braak,Zhang,Chen,Zhong}. These new tools might be useful for treating problems similar to the ones considered here. For example, the new techniques developed in Refs.~\cite{Braak,Zhang,Chen,Zhong} could make it possible to analyze the case of slow-passage LZS interferometry, which would correspond to very small values of $\hbar\omega/\Delta$ and is therefore challenging for us to treat using our numerical approach.

As there is continuing effort to increase the coupling strength in cavity-QED systems with remarkable recent progress and as driving quantum systems with oscillating fields is one of the main tools of quantum control, we expect that the present study combining these two important problems will be relevant to future studies that push the limits of Landau-Zener-Stueckelberg interferometry and cavity QED.

We would like to thank S.~Shevchenko for useful discussions.

\section*{Appendix A: Approximation for Bessel functions obtained from LZS interferometry}

In this appendix we describe an approximation for Bessel functions that arises naturally from analyzing the problem of LZS interferometry.

As mentioned in the main text, the Rabi frequency when strongly driving a $k$-photon resonance with $\hbar\omega\gg\Delta$ is given by
\begin{equation}
\Omega_{\rm Rabi} = \frac{\Delta}{\hbar} J_k\left(\frac{A}{\omega}\right).
\label{Eq:RabiFrequencyAsBesselFunction}
\end{equation}
This expression is expected to be valid for all values of $A/\omega$. There is a well-known approximation for the Bessel function $J_k(x)$ that is valid in the limit $x\gg k$, namely
\begin{equation}
J_k(x) \approx \sqrt{\frac{2}{\pi x}} \cos \left( x - (2k+1) \frac{\pi}{4}\right),
\end{equation}
which gives the approximation
\begin{equation}
\Omega_{\rm Rabi} \approx \frac{\Delta}{\hbar} \sqrt{\frac{2\omega}{\pi A}} \cos \left( \frac{A}{\omega} - (2k+1) \frac{\pi}{4}\right).
\label{Eq:RabiFrequencyWithBesselFunctionApproximation}
\end{equation}

Using a calculation based on the adiabatic-impulse picture, where one constructs the dynamics using well-known expressions for the mixing of probability amplitudes at avoided crossings and the accumulation of phases away from avoided crossings, Ref.~\cite{Ashhab2007} derived an expression for the Rabi frequency that is valid in the limit $\hbar(A-\epsilon)/\Delta\gg 1$ and $\hbar^2\sqrt{A^2-\epsilon^2}\omega/\Delta^2\gg 1$:
\begin{widetext}
\begin{equation}
\Omega_{\rm Rabi} = \frac{\Delta}{\hbar} \sqrt{\frac{2\omega}{\pi \sqrt{A^2-(k\omega)^2}}} \cos \left( \frac{\sqrt{A^2-(k\omega)^2}}{\omega} - k \cos^{-1}\frac{k\omega}{A} - \frac{\pi}{4}\right),
\label{Eq:RabiFrequencyFromAdiabaticImpulseMethod}
\end{equation}
\end{widetext}
where we have used the $k$-photon resonance condition $\epsilon=k\hbar\omega$. This expression differs from the one given in Eq.~(\ref{Eq:RabiFrequencyWithBesselFunctionApproximation}) by the fact that instead of $A$ we now have $\sqrt{A^2-(k\omega)^2}$ and we now have the arccosine function for one of the phases inside the cosine. These differences disappear when we take the limit $A/(k\omega)\gg 1$. However, the expression derived using the adiabatic-impulse method does not require the condition $A/(k\omega)\gg 1$, but rather the less stringent condition $\hbar(A-k\omega)/\Delta\gg 1$. As a result, the expression should remain valid even away from the limit $A\gg(k\omega)$. By comparing Eqs.~(\ref{Eq:RabiFrequencyAsBesselFunction}) and (\ref{Eq:RabiFrequencyFromAdiabaticImpulseMethod}) we can deduce the approximation
\begin{equation}
J_k(x) \approx \sqrt{\frac{2}{\pi \sqrt{x^2-k^2}}} \cos \left( \sqrt{x^2-k^2} - k \cos^{-1}\frac{k}{x} - \frac{\pi}{4} \right).
\end{equation}
This approximation should remain valid as long as $x-k\gg 1$, which is less stringent than the condition $x/k\gg 1$ required for the standard approximation for the Bessel function. Using numerical calculations we have verified that this is indeed the case. In fact, we find that the approximation is very good almost down to the point $x=k$.

We also find that the further approximation where we expand each term inside the cosine function to next-to-leading order in $k/x$, i.e.
\begin{equation}
J_k(x) \approx \sqrt{\frac{2}{\pi \sqrt{x^2-k^2}}} \cos \left( x - k \frac{\pi}{2} - \frac{\pi}{4} + \frac{k^2}{2x} \right),
\end{equation}
is a good approximation almost down to $x=k$ for $k\lesssim 20$.

We have therefore derived a mathematical relation by looking at a physical problem from two different perspectives.


\begin{thebibliography}{99}

\bibitem{Shevchenko} S. N. Shevchenko, S. Ashhab, and F. Nori, Phys. Rep. {\bf 492}, 1 (2010).

\bibitem{Garraway} B. M. Garraway and N. V. Vitanov, Phys. Rev. A {\bf 55}, 4418 (1997).

\bibitem{Shytov} A. V. Shytov, D. A. Ivanov, M. V. Feigel'man, Eur. Phys. J. B {\bf 36}, 263 (2003).

\bibitem{Creffield} C. E. Creffield, Phys. Rev. B {\bf 67}, 165301 (2003).

\bibitem{Shevchenko2005} S. N. Shevchenko, A. S. Kiyko, A. N. Omelyanchouk, W. Krech, Low Temp. Phys. {\bf 31}, 569 (2005).

\bibitem{Ashhab2007} S. Ashhab, J. R. Johansson, A. M. Zagoskin, and F. Nori, Phys. Rev. A {\bf 75}, 063414 (2007).

\bibitem{Son} S.-K. Son, S. Han, and S.-I. Chu, Phys. Rev. A {\bf 79}, 032301 (2009).

\bibitem{Oliver} W. D. Oliver, Y. Yu, J. C. Lee, K. K. Berggren, L. S. Levitov, and T. P. Orlando, Science {\bf 310}, 1653 (2005).

\bibitem{Berns} D. M. Berns, W. D. Oliver, S. O. Valenzuela, A. V. Shytov, K. K. Berggren, L. S. Levitov, and T. P. Orlando, Phys. Rev. Lett. {\bf 97}, 150502 (2006).

\bibitem{Sillanpaa} M. Sillanp\"a\"a, T. Lehtinen, A. Paila, Y. Makhlin, and P. Hakonen, Phys. Rev. Lett. {\bf 96}, 187002 (2006).

\bibitem{Saito2006} S. Saito, T. Meno, M. Ueda, H. Tanaka, K. Semba, and H. Takayanagi, Phys. Rev. Lett. {\bf 96}, 107001 (2006).

\bibitem{Wilson} C. M. Wilson, T. Duty, F. Persson, M. Sandberg, G. Johansson, and P. Delsing, Phys. Rev. Lett. {\bf 98}, 257003 (2007).

\bibitem{Sun} G. Sun, X. Wen, Y. Wang, S. Cong, J. Chen, L. Kang, W. Xu, Y. Yu, S. Han and P. Wu, Appl. Phys. Lett. {\bf 94}, 102502 (2009).

\bibitem{VanDitzhuijzen} C. S. E. van Ditzhuijzen, A. Tauschinsky, and H. B. van Linden van den Heuvell, Phys. Rev. A {\bf 80}, 063407 (2009).

\bibitem{Petta} J. R. Petta, H. Lu, and A. C. Gossard, Science {\bf 327}, 669 (2010).

\bibitem{Childress} L. Childress and J.McIntyre, Phys. Rev. A {\bf 82}, 033839 (2010).

\bibitem{Zhou} J. Zhou, P. Huang, Q. Zhang, Z. Wang, T. Tan, X. Xu, F. Shi, X. Rong, S. Ashhab, and J. Du, Phys. Rev. Lett. {\bf 112}, 010503 (2014).

\bibitem{Silveri} M. P. Silveri, K. S. Kumar, J. Tuorila, J. Li, A. Veps\"al\"ainen, E. V. Thuneberg, and G. S. Paraoanu, New J. Phys. {\bf 17}, 043058 (2015).

\bibitem{Neilinger} P. Neilinger, S. N. Shevchenko, J. Bog\'ar, M. Reh\'ak, G. Oelsner, D. S. Karpov, U. H\"ubner, O. Astafiev, M. Grajcar, and E. Il'ichev, Phys. Rev. B {\bf 94}, 094519 (2016).

\bibitem{Jaynes} E. T. Jaynes and F. W. Cummings, Proc. IEEE {\bf 51}, 89 (1963).

\bibitem{Rabi} I. I. Rabi, Phys. Rev. {\bf 51(8)}, 652 (1937).

\bibitem{WeakCouplingFootnote} A more general condition for the weak-coupling regime is $\lambda\ll \max\{E_{\rm q},\sqrt{E_{\rm q}\hbar\omega}/2\}$, as discussed e.g.~in Ref.~\cite{Ashhab2010}.

\bibitem{Shirley} J. H. Shirley, Phys. Rev. {\bf 138}, B979 (1965).

\bibitem{Nakamura} Y. Nakamura, Y. A. Pashkin, and J. S. Tsai, Phys. Rev. Lett. {\bf 87}, 246601 (2001).

\bibitem{Saito2004} S. Saito, M. Thorwart, H. Tanaka, M. Ueda, H. Nakano, K. Semba, and H. Takayanagi, Phys. Rev. Lett. {\bf 93}, 037001 (2004).

\bibitem{Irish} E. K. Irish, J. Gea-Banacloche, I. Martin, and K. C. Schwab, Phys. Rev. B {\bf 72}, 195410 (2005); E. K. Irish, Phys. Rev. Lett. {\bf 99}, 173601 (2007).

\bibitem{Niemczyk} T. Niemczyk, F. Deppe, H. Huebl, E. P. Menzel, F. Hocke, M. J. Schwarz, J. J. Garcia-Ripoll, D. Zueco, T. H\"ummer, E. Solano, A. Marx, and R. Gross, Nature Phys. {\bf 6}, 772 (2010).

\bibitem{FornDiaz2010} P. Forn-Diaz, J. Lisenfeld, D. Marcos, J. J. Garcia-Ripoll, E. Solano, C. J. P. M. Harmans, and J. E. Mooij, Phys. Rev. Lett. {\bf 105}, 237001 (2010).

\bibitem{Casanova} J. Casanova, G. Romero, I. Lizuain, J. J. Garcia-Ripoll, and E. Solano, Phys. Rev. Lett. {\bf 105}, 263603 (2010).

\bibitem{Wolf} F. A. Wolf, M. Kollar, and D. Braak, Phys. Rev. A {\bf 85}, 053817 (2012).

\bibitem{Yoshihara} F. Yoshihara, T. Fuse, S. Ashhab, K. Kakuyanagi, S. Saito, and K. Semba, Nat. Phys. {\bf 13}, 44 (2017).

\bibitem{YoshiharaSpectra} F. Yoshihara, T. Fuse, S. Ashhab, K. Kakuyanagi, S. Saito, and K. Semba, arXiv:1612.00121.

\bibitem{Ashhab2010} See also S. Ashhab and F. Nori, Phys.~Rev.~A {\bf 81}, 042311 (2010); S. Ashhab, Phys.~Rev.~A {\bf 87}, 013826 (2013); M.-J. Hwang, R. Puebla, and M. B. Plenio, Phys. Rev. Lett. {\bf 115}, 180404 (2015); Z.-J. Ying, M. Liu, H.-G. Luo, H.-Q. Lin, and J. Q. You, Phys. Rev. A {\bf 92}, 053823 (2015).

\bibitem{Deng} C. Deng, J.-L. Orgiazzi, F. Shen, S. Ashhab, and A. Lupascu, Phys. Rev. Lett. {\bf 115}, 133601 (2015); C. Deng, F. Shen, S. Ashhab, and A. Lupascu, Phys. Rev. A {\bf 94}, 032323 (2016).

\bibitem{Lu} See also Z. L\"u and H. Zheng, Phys. Rev. A {\bf 86}, 023831 (2012); Y. Yan, Z. L\"u, and H. Zheng, Phys. Rev. A {\bf 91}, 053834 (2015).

\bibitem{Braak} D. Braak, Phys. Rev. Lett. {\bf 107}, 100401 (2011).

\bibitem{Zhang} Y. Zhang, G. Chen, L. Yu, Q. Liang, L. Q. Liang, and S. Jia, Phys. Rev. A {\bf 83}, 065802 (2011).

\bibitem{Chen} Q. H. Chen, C. Wang, S. He, T. Liu, and K. L. Wang, Phys. Rev. A {\bf 86}, 023822 (2012).

\bibitem{Zhong} H. Zhong, Q. Xie, M. Batchelor, and C. Lee, J. Phys. A: Math. Theor. {\bf 46}, 415302 (2013).

\end{thebibliography}
\end{document}